\documentclass{ws-p8-50x6-00}

\usepackage{latexsym}
\usepackage{amsfonts}
\usepackage{times}

\begin{document}
\newcommand{\h}{H_{\mathrm{eff}}}

\title{Exact Classical Effective Potentials}

\author{J. H. Samson}

\address{Department of Physics,
Loughborough University,
Loughborough,
Leics  LE11 3TU
UK \\ E-mail: j.h.samson@lboro.ac.uk} 


\maketitle 
\begin{center}
	\textbf{Presented at Sixth International Conference on Path Integrals from peV 
	to TeV, Firenze, 26 August 1998}
\end{center}
\abstracts{ A quantum spin system can be modelled by an 
equivalent classical system, with an effective Hamiltonian obtained by 
integrating all non-zero frequency modes out of the path integral.  
The effective Hamiltonian $\h(\{{\mathbf S}_{i}\})$ derived from the 
coherent-state integral is highly singular: the quasiprobability 
density $\exp(-\beta \h)$, a Wigner function, imposes quantisation 
through derivatives of delta functions.  This quasiprobability is the 
distribution of the time-averaged lower symbol of the spin in the 
coherent-state integral.  We relate the quantum Monte Carlo minus-sign 
problem to the non-positivity of this quasiprobability, both 
analytically and by Monte Carlo integration.}

\section{Introduction} \label{intro}
\begin{quote}
	\emph{It is possible, therefore, that a closer study of the relation of classical and quantum 
		theory might involve us in negative probabilities, and so it 
		does.} R P Feynman\cite{Feynman}  
\end{quote}
One may distinguish two broad numerical approaches to quantum 
statistical mechanics.  One can evaluate a path integral by direct 
Monte Carlo methods.  Here the notorious \emph{minus-sign 
problem} (or \emph{phase problem}) often hinders direct evaluation of the 
path integral: this refers to a rapid oscillation of the 
integrand in sign or phase, and the resulting intolerably 
slow convergence of integrals evaluated by random sampling.  Alternatively, one can integrate out the quantum 
fluctuations, leaving an effective Hamiltonian $\h(\mathbf{x})$ with 
\textit{c}-number variables $\mathbf{x}$ to be used in a 
\emph{classical} simulation\cite{GT}.  A similar 
non-positivity arises here if the 
observables described by $\mathbf{x}$ are incompatible: the 
distribution $\exp(-\beta\h(\mathbf{x}))$ is then a Wigner function, 
which is not in general positive-definite.

A free spin $s$ serves here as a useful toy model to relate the path 
integral and the Wigner function\cite{Samson}.  The spin 
coherent-state path integral, a poor starting point for numerical 
simulations, presents the worst case of the sign problem: for 
continuous paths the integrand is a pure Berry phase factor.  The spin 
Wigner function is highly singular, consisting of derivatives of delta 
functions supported on concentric spheres of quantised 
radius\cite{Samson,SW}.  We demonstrate the common origin of these 
non-positivities both analytically and numerically.

\section{Classical effective Hamiltonian}
\label{sec:Heff}
 We consider a system with density matrix 
 $\hat{\rho}\equiv\exp(-\beta\hat{H})$, and assign $c$-number 
 variables $\mathbf{x} \in \mathbb{R}^{N}$ to operators $\mathbf{\hat{x}}$.  
 We define the \emph{Wigner function} $W(\mathbf{x})$ and \emph{classical 
  effective Hamiltonian} $\h(\mathbf{x})$ of these variables as
 \begin{equation}
  	W(\mathbf{x}) \equiv \exp(-\beta \h(\mathbf{x})) = \mathrm{Tr} \left(\hat{\rho} 
  	  	\delta_{N}(\mathbf{x}-\hat{\mathbf{x}})\right),
  	\label{eq:Weyl}
  \end{equation}
where $\delta_{N}$ is the $N$-dimensional delta function.  To remove 
operator-ordering ambiguity, we define
  \begin{equation}
		\delta_{N}(\mathbf{x}-\hat{\mathbf{x}}) \equiv 
		\int\frac{d^{N}\lambda}{(2\pi)^{N}}
		\exp(i\lambda\cdot(\mathbf{x}-\hat{\mathbf{x}})).
    	\label{eq:delta}
    \end{equation}
If the components of $\hat{x}$ do not commute, the delta operator 
need not be positive.

Here we restrict consideration to the spin operator 
$\mathbf{\hat{x}}=\mathbf{\hat{S}}$ in the spin-$s$ representation.  The distribution is the spin 
Wigner function $W_{s}(\mathbf{S})$.  For a single spin with vanishing 
Hamiltonian the trace of Eq.  (\ref{eq:delta}) is easily 
evaluated\cite{Samson,SW} to give derivatives of delta functions 
supported on concentric spheres of quantised radius:
\begin{equation}
	W_{s}(\mathbf{S})  =  \langle\delta_{3}(\mathbf{S}-\hat{\mathbf{S}})\rangle
	= \frac{-1}{2s+1}\sum_{m=-s}^{s}\frac{1}{2\pi 
		S}\frac{d}{dS}\delta(S-m),
		\label{eq:deltaprime}
\end{equation}
%
%
where $S=|\mathbf{S}|$ and $\hat{\mathbf{S}}^{2}=s(s+1)$.  The 
spheres with $m<0$ do not contribute and for \emph{integer} spin there is a 
positive delta function at the origin of weight
\begin{equation}
	\frac{-\delta^{\prime}(S)}{2\pi S(2s+1)}
	= \frac{\delta_{3}(\mathbf{S})}{(2s+1)}.
	\label{eq:deltaeq}
\end{equation}

To motivate this form, we need to verify that the correct marginal 
distributions are obtained.  Integrating $W_{s}$ over a plane 
yields the expected distribution
$\sum_{m=-s}^{s}\delta(\mathbf{e}\cdot\mathbf{S}-m)/(2s+1)$ for the normal 
component of spin\cite{Samson}.  For two coupled spins $1/2$, with Hamiltonian 
$-J\mathbf{S}_{1}\cdot\mathbf{S}_{2}$, a convolution integral
of the two single-spin functions $W_{1/2}$ gives (after some 
manipulations)
\begin{equation}
	W(\mathbf{S}_{1}+\mathbf{S}_{2}) = \frac
	{3e^{\beta J/4}W_{1}(\mathbf{S}_{1}+\mathbf{S}_{2}) +
	e^{-3\beta J/4}W_{0}(\mathbf{S}_{1}+\mathbf{S}_{2})}
	{3e^{\beta J/4} + e^{-3\beta J/4}}.
	\label{eq:W12}
\end{equation}
This is the correct thermal average of triplet and singlet 
distributions, in line with the effective-Hamiltonian interpretation. 
Eq. (\ref{eq:W12}) can be further interpreted as a 
local hidden variable distribution, which must give the correct
correlations for spin measurements on two electrons.  No positive 
definite distribution exists for a singlet state\cite{SW,Bell}, and the 
distribution must therefore be non-definite.

\section{Coherent-state path integrals}
Spin coherent states $|\mathbf{n}\rangle$ for spin $s$ are labelled by 
a unit vector $\mathbf{n}$, such that 
$\mathbf{n\cdot\hat{S}}|\mathbf{n}\rangle = s|\mathbf{n}\rangle$.  To 
relate the Wigner functions just described to the coherent state 
path integral, we split the exponent of Eq. (\ref{eq:delta}) 
into $L$ time slices,
  \begin{equation}
		W_{s}(\mathbf{S}) = \langle\delta_{3}(\mathbf{S}-\hat{\mathbf{S}})\rangle =
		\int\frac{d^{3}\mathbf{\lambda}}{8\pi^{3}}
		e^{i\mathbf{\lambda\cdot S}}
		\left\langle\left(1-i\mathbf{\lambda\cdot\hat{S}}/L+O(L^{-2})\right)^{L}\right\rangle.
    	\label{eq:sliced}
    \end{equation}
We now insert a resolution of unity in the form $\frac{2s+1}{4\pi}\int 
d\mathbf{n}|\mathbf{n}\rangle\langle \mathbf{n}|$ between each 
slice\cite{Klauder}, represent the spin by its \emph{lower} symbol 
$(s+1)\mathbf{n}$\cite{Lieb,DK},
\begin{equation}
	\mathbf{\hat{S}} = \frac{2s+1}{4\pi}\int d\mathbf{n} 
	(s+1)\mathbf{n} |\mathbf{n}\rangle \langle \mathbf{n}| ,
	\label{eq:corr}
\end{equation}
and re-exponentiate.  The lower symbol is used as the paths (here and 
in our Monte Carlo calculation) are only piecewise continuous.  This 
gives
\begin{equation}
	W_{s}(\mathbf{S}) = \lim_{L\rightarrow\infty}W_{s}^{L}(\mathbf{S}),
	\label{eq:limit}
\end{equation}
the limit defined in the sense of a distribution, with 
approximants
  \begin{eqnarray}
& W_{s}^{L}(\mathbf{S}) & = \frac{1}{2s+1}\int\frac{d^{3}\mathbf{\lambda}}{8\pi^{3}} 
e^{i\mathbf{\lambda\cdot S}} \int 
\prod_{i=1}^{L}\left(\frac{2s+1}{4\pi}d\mathbf{n}_{i} 
e^{-i(s+1)\mathbf{\lambda \cdot n}_{i}}\right) \times \nonumber \\ & & \times 
\left\langle\mathbf{n}_{1}\left|\right.\mathbf{n}_{2}\right\rangle 
\left\langle\mathbf{n}_{2}\left|\cdots\right|\mathbf{n}_{L}\right\rangle 
\left\langle\mathbf{n}_{L}\left|\right.\mathbf{n}_{1}\right\rangle 
  \label{eq:WL}\\
& = & \frac{1}{2s+1}\int 
\prod_{i=1}^{L}\left(\frac{2s+1}{4\pi}d\mathbf{n}_{i} \right)
\delta_{3}(\mathbf{S}-(s+1)\mathbf{\bar{n}}) 
\left\langle\mathbf{n}_{1}\left|\right.\mathbf{n}_{2}\right\rangle 
\cdots
\left\langle\mathbf{n}_{L}\left|\right.\mathbf{n}_{1}\right\rangle,
\nonumber
\end{eqnarray}
where $\mathbf{\bar{n}}=\sum_{i=1}^{L}\mathbf{n}_{i}/L$ is the 
time-averaged coherent-state label.  We therefore have a sequence of 
well-defined distributions (\ref{eq:WL}) convergent on the Wigner 
function.  For large $L$ the discretised 
distribution $W_{s}^{L}(\mathbf{S})$ can be shown to be asymptotically a smeared 
Wigner function; for spin $1/2$ we find
\begin{equation}
	W_{1/2}^{L}(\mathbf{S}) \approx \left(\frac{L}{\pi}\right)^{3/2}\int 
	d^{3}\mathbf{x}
	e^{-L(\mathbf{x-S})^{2}} W_{1/2}(\mathbf{S}).
	\label{eq:conv}
\end{equation}


\begin{figure}[t]
\epsfig{file=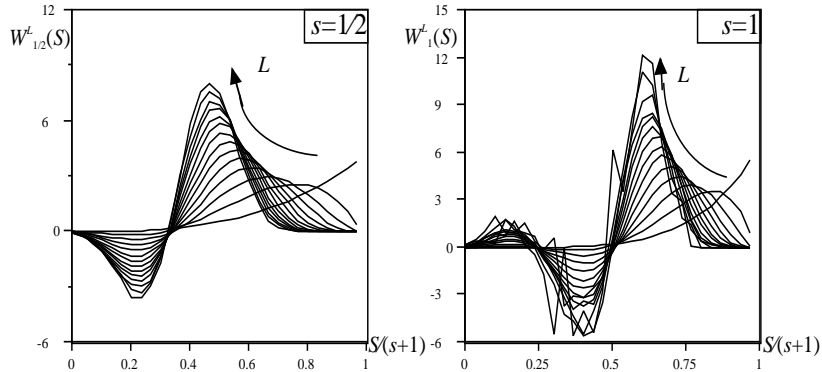, height = 5.5cm, width=11cm}
\caption{Discretised Wigner function for spin $1/2$ and $1$ with $L=2\ldots 15$ time 
slices and $10^{7}$ Monte Carlo steps.
\label{fig:W}}
\end{figure}

In order to verify this result and investigate the importance of the 
sign problem, we have computed the discretised Wigner function 
(\ref{eq:WL}) by direct Monte Carlo sampling of the histogram for 
$\mathbf{\bar{n}}$.  Each path comprises $L$ independent points 
$\mathbf{n}_{i}, i=1\ldots L,$ taken from a uniform distribution on 
the sphere.  Since the product of overlaps in Eq. \ref{eq:WL} is 
complex for $L>2$, a phase problem arises.  For the results shown 
in Fig. \ref{fig:W}, $10^{7}$ independent paths were generated for 
each $L$ from 2 to 15.  Further analysis shows the 
numerical distributions converging to the Wigner function 
(\ref{eq:deltaprime}) according to Eq. (\ref{eq:conv}).  We also see 
how the phase fluctuations lead to numerical instabilities.  The 
correct distribution would be obtained if the the number of Monte 
Carlo steps were taken to infinity before the number of time slices.  
It is clear from the figure that $10^{7}$ steps are insufficient for 
convergence for spin $1$; results for higher spin show still slower 
convergence.

We have thus demonstrated the emergence of a non-positive 
quasiprobability (the spin Wigner function) from the Berry phases of 
the integral.  Numerical calculations are however rapidly 
overwhelmed by the sign problem.

\end{document}